# Photovoltaic Properties of ZnO Nanoparticle Based Solid Polymeric Photoelectrochemical Cells

Manoranjan Ghosh[*], Mima Kurian[†], P. Veerender, N. Padma, D. K. Aswal, S. K. Gupta, J. V. Yakhmi

*Technical Physics Division, Bhabha Atomic Research Centre*
*Trombay, Mumbai – 400085, INDIA*

**Abstract.** In this work, we report optical and photovoltaic properties of ZnO based solid polymeric photoelectrochemical (PEC) cell. In this device construction, film of ZnO nanoparticle (NP) has been sandwiched between ITO and $LiClO_4$-PEO solid electrolyte. ZnO NP has been used as photo-anode and the electrolyte $LiClO_4$ has been dissolved in a solid solution polyethylene oxide that makes the device lightweight and portable. The charge transfer is performed by $Li^+$ and $ClO_4^-$ ions which serve the purpose of a redox couple in a conventional dye sensitize solar cell (DSSC). Short Circuit Current ($J_{SC}$) shows maximum when the ZnO/ITO interface has been illuminated by a UV light ($350 \pm 15$ nm) passing through the transparent ITO glass substrate. $J_{SC}$ and Open Circuit Voltage ($V_{OC}$) of the best performing device are found to be 7 µA/cm$^2$ and 0.24 V respectively under white light of intensity 15 mW/cm$^2$.

## INTRODUCTION

Photoelectrochemical cell (PEC) consisting of conducting polymers and inorganic nanoparticles (NP) have attracted great interest due to their potential applications in developing low-cost and flexible photovoltaic devices.[1] In a conventional PEC, a liquid electrolyte is used for reduction and oxidation reaction to take place for the mechanism of electron exchange. The redox couple used in the electrolyte is $I^-$ and $I_3^-$.[2] However, the liquid electrolyte has limitations such as large weight, shape flexibility, leakage and instability. Solid polymer electrolyte solution prepared by mixing $LiClO_4$ in PVC has been used in a PEC consisting of ITO/$TiO_2$/PVC–LiClO4/Graphite.[3]

Nano-crystals are found suitable for their large surface to bulk ratio giving wide interfacial area for exciton dissociation into free charge carriers and their transportation.[4] ZnO is a large band gap semiconductor like $TiO_2$ which is widely used in PEC and DSSC.[5] It is an excellent absorber and emitter of UV light and displays change in electrical conductivity by UV illumination. Therefore ZnO NPs can be used as photo-anode in a PEC.[6]

A solid solution of PEO and $LiClO_4$ is a familiar candidate for creation of electric double layer and considered as promising battery material.[7] In this work, we report a PEC constructed by layers of ITO/ZnO/PEO–LiClO4/Cu which uses a solid electrolyte solution ($LiClO_4$ in PEO) and a film of ZnO NPs as photo-anode material. The redox couple used in the electrolyte is $Li^+$ and $ClO_4^-$ and copper sheet is used as counter electrode.

Further photovoltaic devices require good Schottky contact with those dissimilar materials. Since p-type ZnO is not very common, use of conductive polymer for realization of better Schottky contact is preferable. In this work an excellent Schottky contact of ZnO with a solid polymer electrolyte (PEO/$LiClO_4$) has been demonstrated.

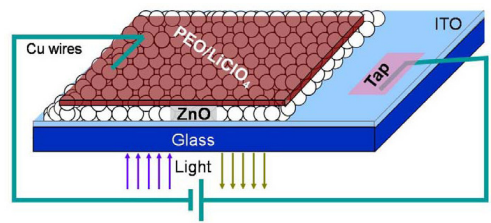

**FIGURE 1:** Illustration of the device fabricated.

[†]Present address: Amity Institute of Nanotechnology, Amity University, Noida-125, U.P.; [*]Email: mghosh@barc.gov.in

## EXPERIMENTAL DETAILS

Figure 1 shows the construction of the fabricated device adopted in this work. ZnO NPs of size ~15 nm have been synthesized by acetate route described elsewhere.[8,9] Films of ZnO NP having thickness ~1.2 and ~0.25 µm are obtained by adding NP dispersion in ethanol on a pre-decided area of a commercially available conducting ITO layer supported by a glass substrate. The ZnO dispersion has been spread over the surface uniformly and dried at room temperature. Equal number of layers was added to obtain films of similar thickness repeatedly. A gel containing 10:1 and 50:1 weight ratio of PEO (poly ethylene oxide, MW 100000) and $LiClO_4$ is deposited on the ZnO layer at room temperature. After 3-4 hours duration, the gel becomes solid and holds the electrical connections tightly as shown in figure 1. The ZnO/ITO interface has been illuminated by a Halogen lamp of intensity 15 mw/cm$^2$ passing through the ITO glass substrate. For IV measurement, positive and negative terminals are connected to the ITO and to the PEO/$LiClO_4$ solid electrolyte respectively through a copper wire. For spectral measurement 1.5 mW/cm$^2$ light of variable wavelength is allowed to pass through the ITO-Glass substrate from Xenon arc lamp fixed in a Spectrofluorimeter (Jobin Yvon Fluuomax 3). $J_{SC}$, $V_{OC}$ and IV measurement have been performed by Keithley, 6487 Voltage Source/Pico-ammeter.

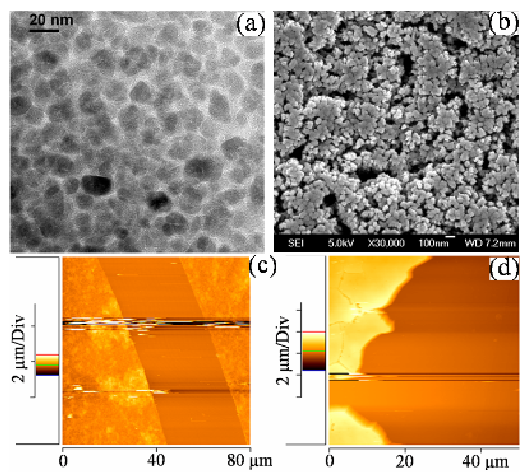

**FIGURE 2.** (a) TEM image of collection of nanoparticle of dia~15 nm. (b) The nanoparticle film when coated on substrate shows agglomerated cluster of size 40 nm. AFM image of scratched film gives film thickness (c) 0.2-0.3 µm and (d) 1.2 µm.

TEM image collected by JEOL HR-TEM shows that average size of the NP is ~15 nm [figure 2 (a)]. When the dispersion of such NPs is deposited on a substrate, agglomeration takes place and the average grain size becomes 40–50 nm [SEM image in figure 2 (b)]. The film thicknesses measured at the step of ZnO/ITO of a scratched film from the AFM (Model-CP2, VEECO) images are found to be 0.2-0.3 µm [figure 2(c)] and 1.2 µm [figure 2(d)].

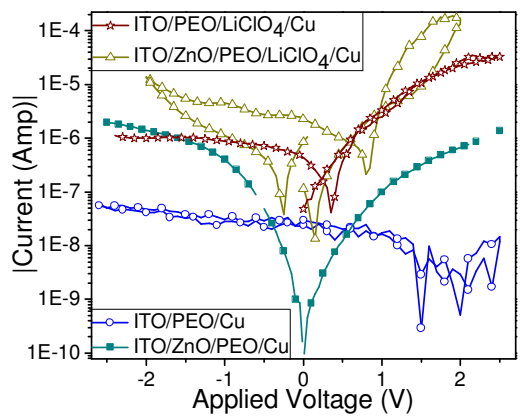

**FIGURE 3.** I-V characteristics in dark mode for various junctions as indicated on the graph.

## RESULTS

### Role of Various Components of the Device

The electrical behavior of the device depends on the individual junction created at the interface of different layers. The ITO layer behaves like traditional conductors below their plasmon frequency and can be treated as metal for the low frequency electrical signal. ZnO is inherently an n-type semiconductor acting as a source of photo-generated carriers and PEO is used as solvent for $LiClO_4$. There are three junctions involved at the interface of ITO/ZnO, ZnO/*PEO-LiClO$_4$* and ITO/*PEO-LiClO$_4$* if accidental contact happens due to the porous nature of the ZnO film.

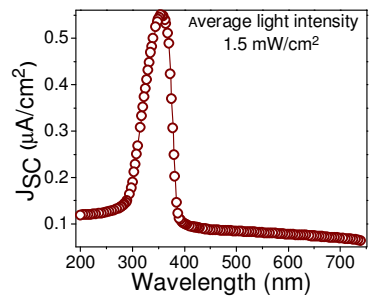

**FIGURE 4.** The spectral response of the device shows a maximum when illuminated by light of wavelength around 350±15 nm.

**TABLE 1.** Short Circuit Current ($J_{SC}$) and Open Circuit Voltage ($V_{OC}$) in Dark and light for the Devices Fabricated.

| Description of the Device | | Open Circuit Voltage ($V_{OC}$) | | Short Circuit Current ($J_{SC}$) | |
|---|---|---|---|---|---|
| PEO:LiClO$_4$ (wt. ratio) | ZnO layer thickness (μm) | Dark (V) | Light of intensity 15 mw/cm$^2$ (V) | Dark (μA/cm$^2$) | Light of intensity 15 mw/cm$^2$ (μA/cm$^2$) |
| 50:1 | 1.2 | 0.065 | 0.325 | 0.04 | 3.5 |
| 10:1 | 0.3 | 0.003 | 0.240 | 0.03 | 7 |

To explore the different contact behavior, the dark I-V characteristics of devices fabricated in the absence of i) polymeric electrolyte and ii) ZnO layer are investigated. As shown in figure 3, the device made of ITO/PEO/Cu shows very low current. When ZnO layer is introduced, the I-V curve shows symmetric behavior. On the other hand ITO/*PEO-LiClO$_4$*/Cu junction exhibits rectification of about 10 at 1.5 V. The rectification increases up to 30 when the complete device ITO/ZnO/*PEO-LiClO$_4$*/Cu is considered. Note that the graphs are plotted in semi-log scale to blow up the low current regime. The above observation shows that nonlinearity in the I-V characteristic of the device mainly arises due to the semiconductor (ZnO)-electrolyte (PEO-LiClO$_4$) junction. Only use of polymer (in the absence of *LiClO$_4$*) does not show much rectification. Electrolyte in the polymer acts as a source of electrical charge and enhances device current (*PEO:LiClO$_4$*=10:1) by three orders of magnitude compare to the case when no electrolyte is used.

## Spectral Response

We have worked out the suitable range of optical wavelength over which the device can perform best. In the range of 320 – 375 nm, ZnO is a good absorbing material but ITO-Glass substrate is transparent above 320 nm. Therefore, UV light of wavelength in the range 320 – 375 nm can easily pass through the ITO glass substrate but will be absorbed fully by the thick ZnO layer. In figure 4 we plot the $J_{SC}$ after illuminating the device over a range of spectrum. It can be easily seen that the current through the device is maximum when illuminated in the range of 350±15 nm. This is a property of wide band gap ZnO due to which excess carrier generation (EHPs) takes place only after excitation by a light of energy more than the band gap. The fall in the photocurrent below 330 nm is due to the absorption of light by the ITO glass substrate.

## Photovoltaic Properties

$J_{SC}$ and $V_{OC}$ have been measured on 0.2 cm$^2$ areas for two devices having different LiClO$_4$ conc. and ZnO layer thickness. It can be seen from TABLE 1 that material combination used in this work exhibits higher $J_{SC}$ and $V_{OC}$ compared to the earlier reported TiO$_2$ based PEC which uses solid polymer electrolyte[3]. $V_{OC}$ is found to increase when ZnO layer thickness increases and LiClO$_4$ conc. decreases. In contrary, large $J_{SC}$ is recorded at higher LiClO$_4$ conc. and lower ZnO thickness. Currently, we are optimizing the ZnO layer thickness and LiClO$_4$ conc. as well as investigating the current transport mechanism of the device. The type of PEC reported here shows large dark $V_{OC}$ (0.2 V) immediately after the fabrication. After week long shortening of the two terminals, dark $V_{OC}$ decreases and becomes negligible. However, dark $J_{SC}$ values show negligible variations due to shortening.

## CONCLUSIONS

In conclusion, a ZnO based PEC has been fabricated by low cost solution method which uses PEO/LiClO$_4$ solid polymer electrolyte. ZnO NP forms an excellent Schottky contact with the polymer electrolyte. UV light of wavelength 350±15 nm produces highest photocurrent in the device. The device show improved efficiency compared to the TiO$_2$ based PEC of its kind but it is lower than the PEC making use of liquid electrolyte mainly due to its low $J_{SC}$ value.